\documentclass{moriond}

\def\be{\begin{equation}}
\def\ee{\end{equation}}
\def\bea{\begin{eqnarray}}
\def\eea{\end{eqnarray}}

\newcommand{\Photo}{}

\begin{document}
\vspace*{4cm}
\title{The ESSnuSB Experiment}

\author{Monojit Ghosh on behalf of the ESSnuSB Collaboration}

\address{Center of Excellence for Advanced Materials and Sensing Devices, Ru{\dj}er Bo\v{s}kovi\'c Institute, 10000 Zagreb, Croatia}

\maketitle

\abstracts{
In this proceedings, we will describe the physics program of the ESSnuSBplus, phase-I of the ESSnuSB project. ESSnuSB is a future long-baseline neutrino oscillation experiment in Europe which aims to measure $\delta_{\rm CP}$ at the second oscillation maximum with with unprecedented precision. Apart from studying the beam based physics, the large far detector is also capable of studying various other physics cases involving solar, atmopsheric and supernova neutrinos. Under the ESSnuSBplus project, there will be a low energy monitored beam and a low energy nuSTORM facility for the measurement of cross-section.
}

\section{Introduction}

While the current generation of the long-baseline neutrino experiments T2K and NO$\nu$A \cite{T2K:2025wet} are hinting towards the true nature of the neutrino mass ordering, true octant of the atmospheric mixing angle $\theta_{23}$ and CP violation in the leptonic sector, the goal of the future generation experiments T2HK \cite{Hyper-Kamiokande:2018ofw} and DUNE \cite{DUNE:2020ypp}, will be to establish these hints in a firm footing. The main objective of the ESSnuSB experiment \cite{Alekou:2022emd} will be to improve the measurement of the leptonic CP violating phase $\delta_{\rm CP}$ to an unprecedented precision in order to understand the correct theory for physics responsible for this CPV. Apart from the measurement of $\delta_{\rm CP}$, the large far detector is also capable of detecting neutrinos from Sun, Earth's atmosphere and future supernova explosion. The ongoing Horion-Europe ESSnuSBplus project is a collaboration gathering 13 countries and 23 institutes. Within this project, the low energy monitored beam and the low energy nuSTORM facilities will measure the neutrino-nucleus cross-section in the sub-GeV energy region to reduce the systematic uncertainties. 

This proceedings is organized as follows. In section 2, we will discuss briefly the physics potential of the ESSnuSB experimet. In section 3, we will introduce the physics goals within the ESSnuSBplus setup. Finally we will summarize and conclude.

\section{The ESSnuSB}
\label{subsec:prod}

In the ESSnuSB experiment, neutrinos will be produced at the ESS facility in Lund Sweden. It will use the 5 MW ESS-linac proton beam with 2.5 GeV proton energy. These neutrinos will be detected at the Zinkgruvan mine is Sweden which is located at a distance of 360 km from Lund. The far detector will consist of a water Cerenkov detector with a fiducial volume of 538 kt. 

\begin{figure}
\centering
\includegraphics[width=0.7\linewidth]{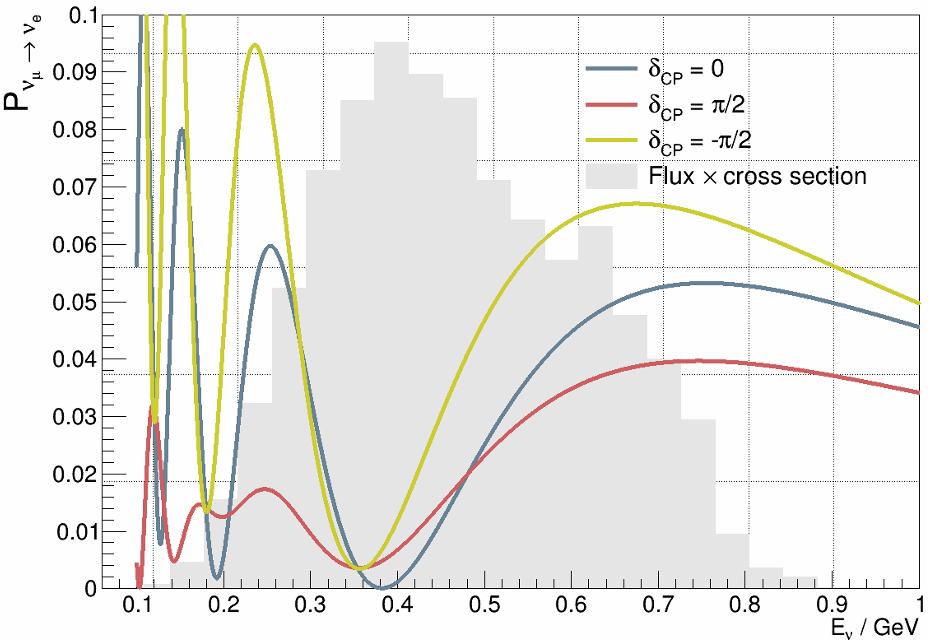}
\caption[]{Probability and Flux.}
\label{fig1}
\end{figure}

Fig.~\ref{fig1}, shows the appearance channel probability as a function of neutrino energy. The shaded histogram shows the flux folded with cross-section, demonstrating the energy region to which ESSnuSB is sensitive to. From this figure we understand that, the separation between the CP violating and the CP conserving curves are large at the second oscillation maximum as compared to that at the first oscillation maximum. It also shows that ESSnuSB covers a part of the first oscillation maximum and a part of the second oscillation maximum. 

\begin{figure}[h]
\includegraphics[width=0.48\linewidth]{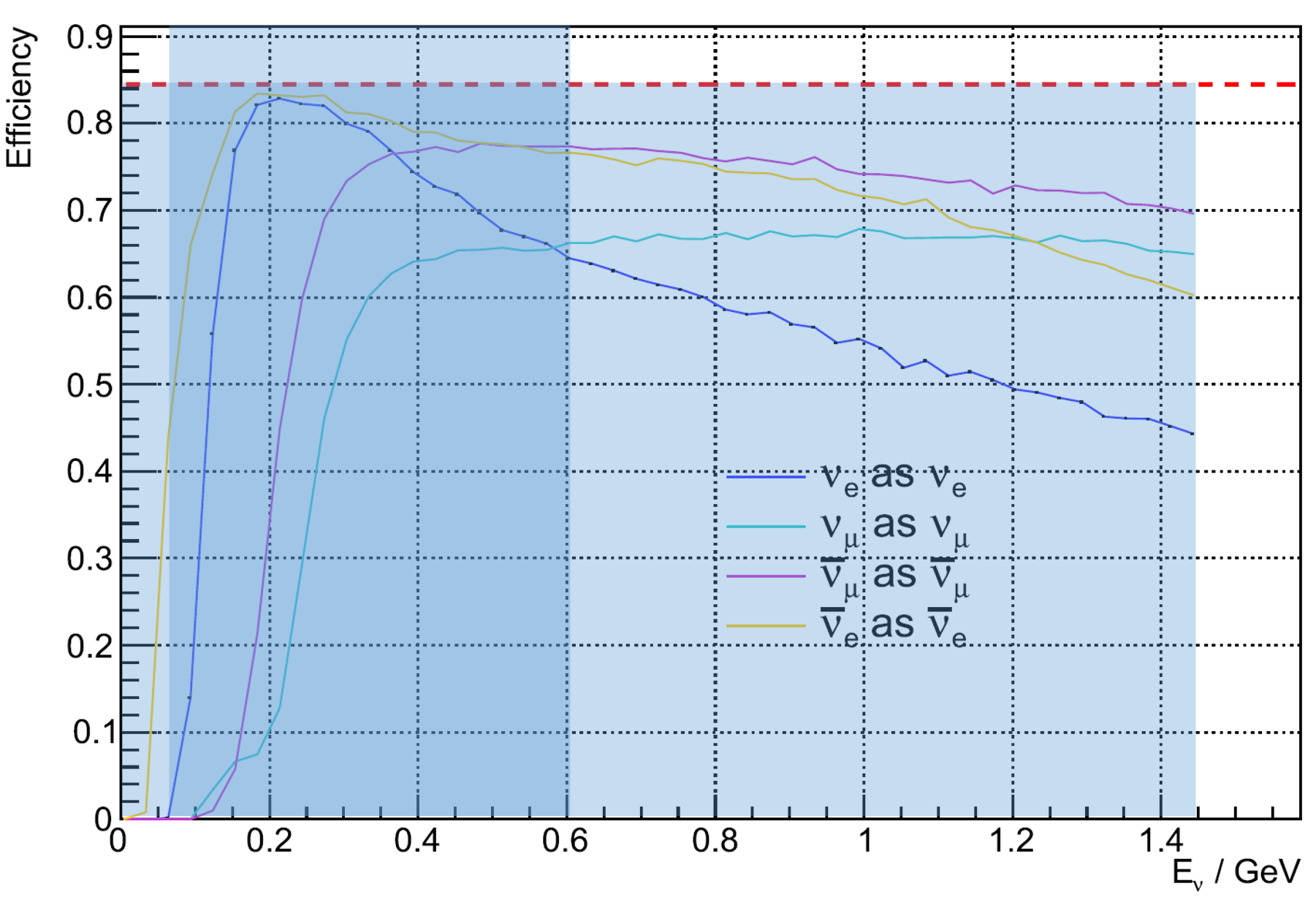}
\includegraphics[width=0.53\linewidth]{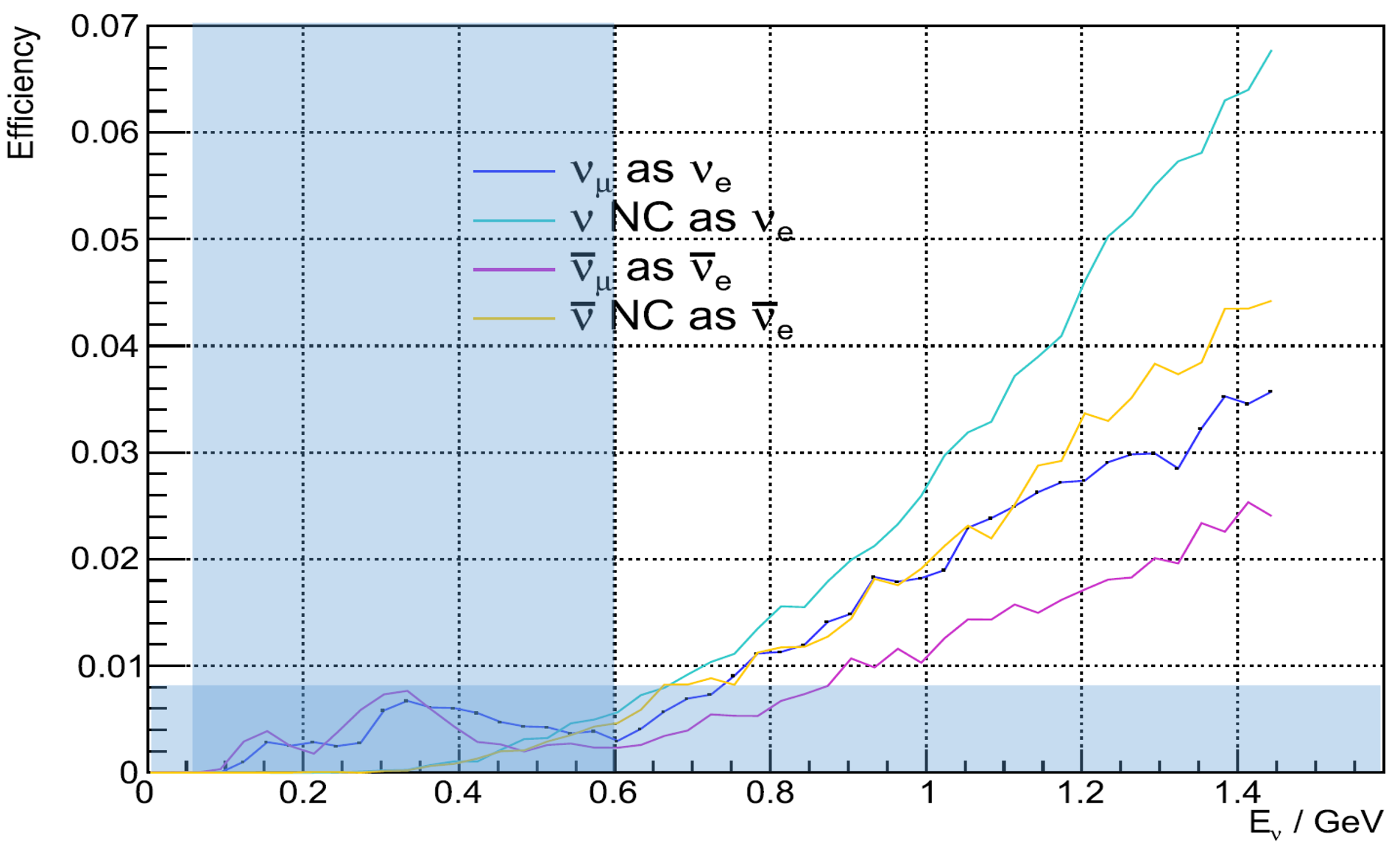}
\caption[]{Detector efficiency for signal (left panel) and background (right panel).}
\label{fig2}
\end{figure}

Fig.\ref{fig2} shows the efficiencies of the ESnuSB far detector optimized for the beam based physics. The left panel shows the efficiency for the signal events and the right panel shows the efficiency for the background events. From the left panel we see that ESSnuSB has around 85\% efficiency for the acceptance of the $\nu_e$ signal events and more than 99\% efficiency for the rejection of the background events in the energy region where the flux folded by cross-section peaks. 

\begin{figure}[h]
\includegraphics[width=0.5\linewidth]{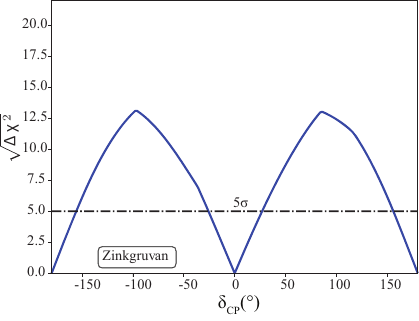}
\includegraphics[width=0.5\linewidth]{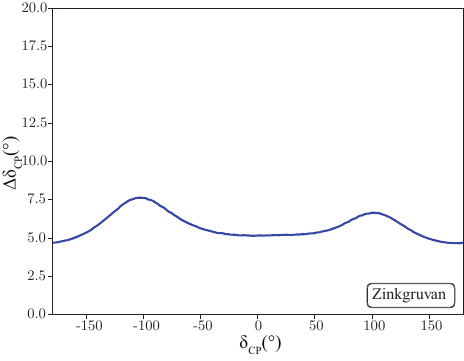}
\caption[]{CP violation (left panel) and CP precision (right panel) sensitivity.}
\label{fig3}
\end{figure}

Fig.~\ref{fig3} shows the CP sensitivity of the ESSnuSB experiment as a function of true $\delta_{\rm CP}$. The left panel is for CP violation and the right panel is for CP precision. From the figure wee see that one can achieve 12 $\sigma$ CP violation sensitivity for the maximal values of $\delta_{\rm CP}$ and less than $7^\circ$ CP precision sensitivity for the all the values of $\delta_{\rm CP}$. 

Apart from the CP sensitivity, the ESSnuSB FD is very useful for studying various new physics scenarios based on neutrino beam \cite{ESSnuSB:2025shd,ESSnuSB:2024yji,ESSnuSB:2023lbg} and also various non-beam based physics scenarios including atmospheric, solar, supernova neutrinos and proton decay \cite{ESSnuSB:2026dar,ESSnuSB:2025vsf,ESSnuSB:2024wet}. 

\section{The ESSnuSBplus}\label{subsec:prod}

\begin{figure}[h]
\centering
\includegraphics[width=1.0\linewidth]{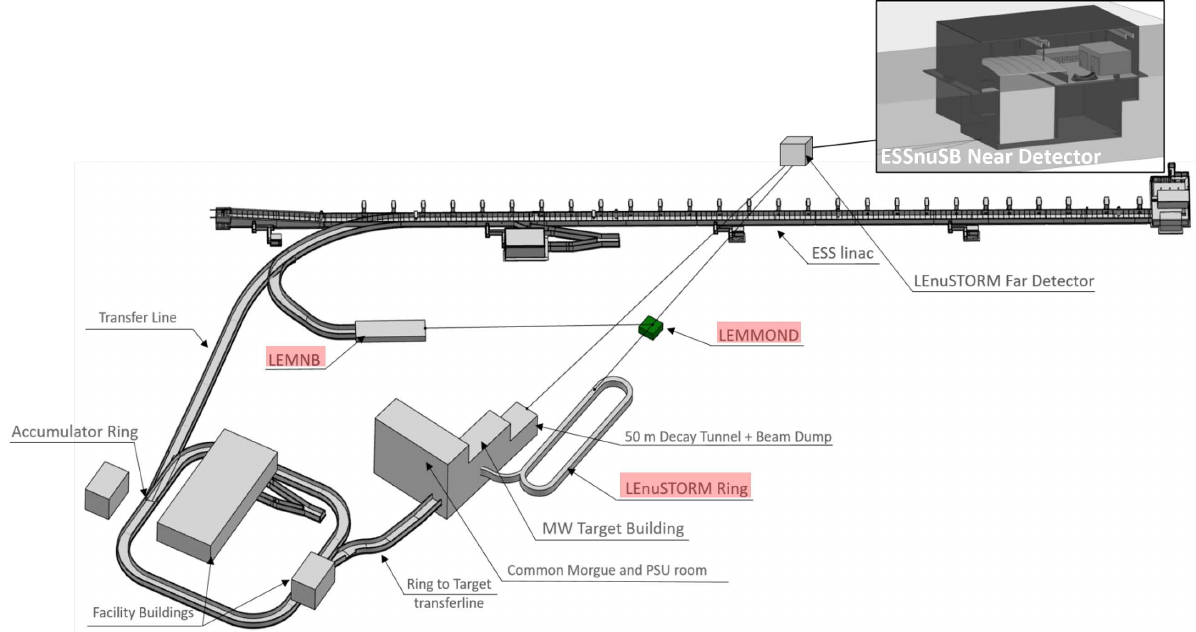}
\caption[]{The ESSnuSB+ setup}
\label{fig4}
\end{figure}

A preliminary layout of the ESSnuSBplus facility has been shown in Fig.~\ref{fig4}. As mentioned earlier, the main ESS linac in Lund will deliver a 5 MW proton beam with 2.5 GeV proton energy and 2.8 ms pulse. Then the accumulator ring will reduce the pulse length to 1.2 $\mu$s in order to minimize atmospheric background. These protons will collide with a target to produce a intense beam of muon neutrinos via pion decay. These neutrinos will pass through a near detector located at 250 m (END) and then they will finally reach the far detector (FD) consisting of 540 kt ultra-pure water, located at a distance of 360 km at Zingruvan. In addition, this facility will also have a low energy monitored neutrino beam (LEMNB) which is inspired by the original ENUBET \footnote{to be implemented in nuSCOPE \cite{Acerbi:2025wzo}.} idea project at CERN \cite{ENUBET:2025qhg}. This will be an instrumented decay pipe to measure neutrino flux to 1\% uncertainty. These  neutrinos will be detected at another near detector called LEMMOND. Moreover, for the measurement of cross-section, this facility will also have a low energy neutrino source from muon decay (LEnuSTORM) inspired by the original nuSTORM idea \cite{nuSTORM:2025tph}. These neutrinos will be detected in LEMMOND and END detectors. 

Apart from the cross-section measurement, the the LEMNB and the LEnuSTORM facilities (and also the main ESS beam) provide opportunities to study sterile neutrinos and other short baseline physics scenarios for example non-standard interactions, non-unitarity etc at the near detectors LEMMOND and END.

\section*{Acknowledgments}

The work in part funded by (i) Ministry of Science Education and Youth of the Republic of Croatia Grant No. PK.1.1.10.0002, (ii) SNSF and HRZZ under grant MAPS IZ11Z0$\_$230193 and (iv) European Union and European Union under the NextGenerationEU Programme. Views and opinions expressed are however those of the author(s) only and do not necessarily reflect those of the European Union. Neither the European Union nor the granting authority can be held responsible for them.

\section*{References}
\bibliography{moriond}

@article{T2K:2025wet,
    author = "Abubakar, S. and others",
    collaboration = "T2K, NOvA",
    title = "{Joint neutrino oscillation analysis from the T2K and NOvA experiments}",
    eprint = "2510.19888",
    archivePrefix = "arXiv",
    primaryClass = "hep-ex",
    reportNumber = "FERMILAB-PUB-25-0132-PPD",
    doi = "10.1038/s41586-025-09599-3",
    journal = "Nature",
    volume = "646",
    number = "8086",
    pages = "818--824",
    year = "2025"
}

@article{Hyper-Kamiokande:2018ofw,
    author = "Abe, K. and others",
    collaboration = "Hyper-Kamiokande",
    title = "{Hyper-Kamiokande Design Report}",
    eprint = "1805.04163",
    archivePrefix = "arXiv",
    primaryClass = "physics.ins-det",
    month = "5",
    year = "2018"
}

@article{DUNE:2020ypp,
    author = "Abi, Babak and others",
    collaboration = "DUNE",
    title = "{Deep Underground Neutrino Experiment (DUNE), Far Detector Technical Design Report, Volume II: DUNE Physics}",
    eprint = "2002.03005",
    archivePrefix = "arXiv",
    primaryClass = "hep-ex",
    reportNumber = "FERMILAB-PUB-20-025-ND, FERMILAB-DESIGN-2020-02",
    doi = "10.2172/1599307",
    month = "2",
    year = "2020"
}

@article{Alekou:2022emd,
    author = "Alekou, A. and others",
    title = "{The European Spallation Source neutrino super-beam conceptual design report}",
    eprint = "2206.01208",
    archivePrefix = "arXiv",
    primaryClass = "hep-ex",
    doi = "10.1140/epjs/s11734-022-00664-w",
    journal = "Eur. Phys. J. ST",
    volume = "231",
    number = "21",
    pages = "3779--3955",
    year = "2022",
    note = "[Erratum: Eur.Phys.J.ST 232, 15--16 (2023)]"
}

@article{ESSnuSB:2026dar,
    author = "Aguilar, J. and others",
    collaboration = "ESSnuSB",
    title = "{Complementarity between atmospheric and super-beam neutrinos at ESSnuSB}",
    eprint = "2603.02836",
    archivePrefix = "arXiv",
    primaryClass = "hep-ex",
    month = "3",
    year = "2026"
}

@article{ESSnuSB:2025vsf,
    author = "Aguilar, J. and others",
    collaboration = "ESSnuSB",
    title = "{Searching non-standard interactions with atmospheric neutrinos at ESSnuSB}",
    eprint = "2508.18103",
    archivePrefix = "arXiv",
    primaryClass = "hep-ex",
    doi = "10.1007/JHEP05(2026)109",
    journal = "JHEP",
    volume = "05",
    pages = "109",
    year = "2026"
}

@article{ESSnuSB:2025shd,
    author = "Aguilar, J. and others",
    collaboration = "ESSnuSB",
    title = "{Probing long-range forces in neutrino oscillations at the ESSnuSB experiment}",
    eprint = "2504.10480",
    archivePrefix = "arXiv",
    primaryClass = "hep-ph",
    doi = "10.1007/JHEP07(2025)186",
    journal = "JHEP",
    volume = "07",
    pages = "186",
    year = "2025"
}

@article{ESSnuSB:2024wet,
    author = "Aguilar, J. and others",
    collaboration = "ESSnuSB",
    title = "{Exploring atmospheric neutrino oscillations at ESSnuSB}",
    eprint = "2407.21663",
    archivePrefix = "arXiv",
    primaryClass = "hep-ex",
    doi = "10.1007/JHEP10(2024)187",
    journal = "JHEP",
    volume = "10",
    pages = "187",
    year = "2024"
}

@article{ESSnuSB:2024yji,
    author = "Aguilar, J. and others",
    collaboration = "ESSnuSB",
    title = "{Decoherence in neutrino oscillation at the ESSnuSB experiment}",
    eprint = "2404.17559",
    archivePrefix = "arXiv",
    primaryClass = "hep-ex",
    doi = "10.1007/JHEP08(2024)063",
    journal = "JHEP",
    volume = "08",
    pages = "063",
    year = "2024"
}

@article{ESSnuSB:2023lbg,
    author = "Aguilar, J. and others",
    collaboration = "ESSnuSB",
    title = "{Study of nonstandard interactions mediated by a scalar field at the ESSnuSB experiment}",
    eprint = "2310.10749",
    archivePrefix = "arXiv",
    primaryClass = "hep-ex",
    doi = "10.1103/PhysRevD.109.115010",
    journal = "Phys. Rev. D",
    volume = "109",
    number = "11",
    pages = "115010",
    year = "2024"
}

@article{Acerbi:2025wzo,
    author = "Acerbi, F. and others",
    title = "{nuSCOPE: A short-baseline neutrino beam at CERN for high-precision cross-section measurements}",
    eprint = "2503.21589",
    archivePrefix = "arXiv",
    primaryClass = "hep-ex",
    month = "3",
    year = "2025"
}

@inproceedings{ENUBET:2025qhg,
    author = "Hali{\'c}, L. and others",
    collaboration = "ENUBET",
    title = "{The ENUBET monitored neutrino beam and its implementation at CERN}",
    booktitle = "{25th International Workshop on Neutrinos from Accelerators}",
    eprint = "2501.04531",
    archivePrefix = "arXiv",
    primaryClass = "hep-ex",
    month = "1",
    year = "2025"
}

@inproceedings{nuSTORM:2025tph,
    author = "Ruso, L. Alvarez and others",
    collaboration = "nuSTORM",
    title = "{Neutrinos from Stored Muons (nuSTORM)}",
    eprint = "2505.06137",
    archivePrefix = "arXiv",
    primaryClass = "hep-ex",
    month = "5",
    year = "2025"
}


\end{document}